\documentclass[twocolumn]{aastex61}

\begin{document}

\title{An efficient statistical method to compute molecular
  collisional rate coefficients}

\author{J\'er\^ome Loreau}
\email{jloreau@ulb.ac.be}
         \affil{LOMC - UMR 6294,
Normandie Universit\'e, Universit\'e du Havre and CNRS, 25 rue Philippe Lebon, BP 1123 - 76 063 Le Havre cedex, France}
        \affil{Service de Chimie Quantique et Photophysique, Universit\'e libre de Bruxelles (ULB) CP 160/09, 1050 Brussels, Belgium}

 \author{Fran\c{c}ois Lique}
\email{francois.lique@univ-lehavre.fr}
         \affil{LOMC - UMR 6294,
Normandie Universit\'e, Universit\'e du Havre and CNRS, 25 rue Philippe Lebon, BP 1123 - 76 063 Le Havre cedex, France}

 \author{Alexandre Faure}
 \email{alexandre.faure@univ-grenoble-alpes.fr}
        \affil{UJF-Grenoble 1/CNRS, Institut de Plan\'etologie et d'Astrophysique de Grenoble (IPAG) UMR 5274, Grenoble F-38041, France}

\begin{abstract}

Our knowledge about the ``cold'' Universe often relies on molecular
spectra. A general property of such spectra is that the energy level
populations are rarely at local thermodynamic equilibrium. Solving the
radiative transfer thus requires the availability of collisional rate
coefficients with the main colliding partners over the temperature
range $\sim$ 10--1000~K. These rate coefficients are notoriously
difficult to measure and expensive to compute. In particular, very few
reliable collisional data exist for collisions involving reactive
radicals or ions.  In this Letter, we explore the use of a fast
quantum statistical method to determine molecular collisional excitation rate
coefficients. The method is benchmarked against accurate (but costly)
close-coupling calculations. For collisions proceeding through the
formation of a strongly-bound complex, the method is found to be
highly satisfactory up to room temperature. Its accuracy decreases
with the potential well depth and with increasing temperature, as
expected. This new method opens the way to the determination of
accurate inelastic collisional data involving key reactive species such as H$_3^+$, H$_2$O$^+$, and H$_3$O$^+$ for which exact quantum calculations are currently not feasible.

\end{abstract}


\section{Introduction}

A general property of astronomical molecular spectra is that the
populations of the energy levels are rarely at local thermodynamical
equilibrium (LTE). 
Indeed, in space, the density is usually such that the frequency of
collisions is neither negligible nor large enough to maintain
LTE. Deviations from LTE, including strong deviations like population
inversions, are thus the rule rather than the exception. In such
conditions, interpreting a spectra requires to solve simultaneously
the radiative transfer equation and a set of statistical equilibrium
equations for the molecular energy levels. Solving the statistical
equilibrium in turn necessitates the availability of the energy
levels, the state-to-state rates for spontaneous radiative decay and
the state-to-state rate coefficients for collisional
(de)excitation. The collisional (inelastic) rate coefficients are
extremely difficult to measure in the laboratory and radiative
transfer models rely almost exclusively on theoretical estimates.

Nowadays theoretical data can be obtained for neutral - closed shell -
diatomic and polyatomic molecules colliding with atoms or light
diatomic perturbers like H$_2$ at the full quantum time-independent
close-coupling level of theory. In addition, accuracy can be inferred
by comparing calculations directly with state-to-state experimental
data thanks to recent developments in double resonance
\citep{mertens17} and crossed-beam techniques
\citep{vogels15,chefdeville15}. Such comparisons have demonstrated the
high-accuracy of modern computations based on state-of-the-art
potential energy surfaces. For (polyatomic) reactive radicals
and ions or polyatomic projectiles, however, the numerical solution of
the coupled equations becomes impractical due to the memory and CPU
requirements\footnote{Memory and CPU costs scale as the square and the
  cube of the total number of channels, respectively. The
  close-coupling approach becomes prohibitive when the number of
  coupled-channels exceed $\sim$ 10,000.}. The problem is particularly
acute for systems with deep potential wells because an excessively
large number of closed channels is needed to achieve convergence. As a
result, for collisions where the intermediate complex is a stable
molecule or ion, close-coupling calculations have been restricted to
the very simplest atom-diatom systems, see
e.g. \cite{stoecklin15,Bulut:15} for OH$^+$+H.

Various approximative methods however exist to treat the most
demanding systems. A standard approach to reduce the number of coupled
channels is to use angular momentum {\it decoupling} approximations
such as the coupled-states and the infinite-order-sudden methods
\citep{kouri79}. Such approximations become however reliable only at
high collision energies with respect to the potential well and/or to
the rotational spacings, which in practice restricts their
applicability to temperatures above, typically, 300~K. Quantum
time-dependent (wave packet) methods suffer similar limitation,
although recent progress has been made with the Multi Configuration
Time Dependent Hartree (MCTDH) method
\citep{ndengue17}. Semi-classical and quasi-classical methods offer
another alternative, their common feature being that at least one
degree of freedom in the collisional system is treated
classically. The good agreement observed between such methods and
full-quantum calculations is encouraging \citep{faure16,semenov17} but
there are still intrinsic limitations, especially when resonances or
interferences are important.

In the present Letter, we explore another class of approximation to treat purely inelastic collisions: the
statistical method. The statistical approach was initially developed
for chemical reactions \citep[see][and references
  therein]{Pechukas1966, Park2007, Gonzalez-Lezana2007, dagdigian-rev}. Statistical models are expected to be
accurate for reactive or inelastic collisions proceeding through
formation and decay of a strongly-bound collision complex. Although
the basic hypothesis of statistical theories are similar, the
different models introduce various dynamical constraints. Recent
examples include the treatment of the reactive collision H$_3^+$ + H$_2$ by \cite{Park2007}, later extended by  \cite{Gomez-Carrasco2012}, or the study of rotationally inelastic
collisions of CH with H$_2$ \citep{dagdigian16} and H
\citep{dagdigian17} using the quantum statistical method of
Manolopoulos and co-workers \citep{rackham01}. Here we employ a simpler
approach inspired by the statistical adiabatic channel theory of
\cite{quack74,quack75}. The method is checked against full quantum
time-independent calculations for a series of benchmark systems,
including ionic and neutral targets, and for a large range of
temperatures. In Section 2, we describe the statistical approach and
the determination of collisional rate coefficients.  In Section 3, we
compare the results obtained with this method to exact quantum
calculations. A discussion on the possible use of our model concludes
this letter.

\section{Theoretical methods}

The close-coupling method is based on an expansion of the total
wavefunction $\Psi$ into an angular basis set $\{ \vert
\alpha\rangle\}$:
\begin{equation}\label{eq_wf}
\Psi = R^{-1} \sum_{\alpha} \vert \alpha \rangle \chi_\alpha (R)
\end{equation}
where $\chi_\alpha (R)$ are radial functions describing the nuclear motion and the form of the angular functions depend on the type of collision \citep{Flower2007}.
Integrating the Schr\"odinger equation over the angular variables leads to the set of coupled differential radial equations
\begin{equation}\label{eq_CC}
-\frac{\hbar^2}{2\mu}\frac{d^2 \chi_\alpha }{dR^2}
+ \sum_{\alpha^\prime}\langle \alpha^\prime\vert H_{\mathrm{int}} + V + \frac{{\bf L}^2}{2 \mu R^2} \vert \alpha\rangle \chi_\alpha = E \chi_\alpha
\end{equation}
to be solved for a given collision energy $E$ and for each value of
the total angular momentum $J$. In this equation, $H_{\mathrm{int}}$
is the Hamiltonian describing the internal ro-vibrational motion of
the molecule(s), $V$ denotes the potential energy surface of the
system, and ${\bf L}$ is the angular momentum describing the relative
motion of the two colliding partners.  The total angular momentum
${\bf J = j_1 + j_2 + L}$, where ${\bf j_1}$ and ${\bf j_2}$ are the
angular momenta of the colliding molecules (${\bf j_2}=0$ for
atom-molecule collisions), can be used to conveniently reformulate the
coupled equations. The total angular momentum is conserved during a
collision, and the coupled equations become block-diagonal in $J$.
Solving these equations with appropriate boundary conditions gives
access to the collision matrix $S(E,J)$ and the inelastic cross
sections are obtained by summing the contributions of all values of
$J$ until convergence is reached.

\begin{figure}
    \epsscale{1.1}
\centering
\label{fig_adiabats}
    \plotone{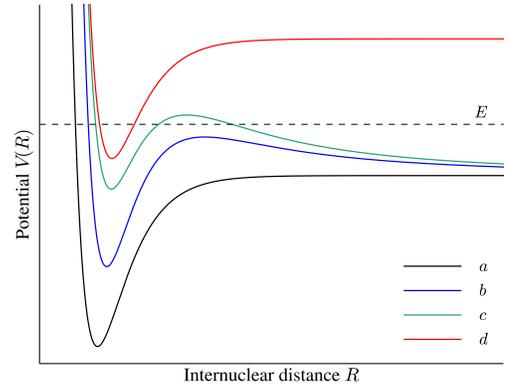}
    \caption{Adiabatic potential energy curves. At the energy shown, channels $a$ and $b$ are open, while channels $c$ and $d$ are closed.}
    \end{figure}

In our approach, we diagonalize the Hamiltonian excluding the nuclear
kinetic term (i.e., only the second term of eq. (\ref{eq_CC})) for
each value of $J$, which results in a set of adiabatic curves. The
diagonalization was performed with the {\small MOLSCAT} program
\citep{molscat} for a grid of internuclear distances in the range
$3-50$~$a_0$. 
It is important to note that the present approach requires an accurate {\it ab initio} potential energy surface for the collisional complex.
Each adiabatic curve can be associated asymptotically
with an internal state of the colliding molecules, and the number of
adiabatic curves is determined by the values that the quantum number
$L$ can take according to the standard angular momentum coupling
rules. Examples of adiabatic channels are represented in
Fig. \ref{fig_adiabats}. The basic assumption is that if the kinetic
energy is larger than the centrifugal barrier in the entrance channel,
an intermediate molecular complex can be formed and an inelastic
collision can take place. The probability of inelastic transition is
determined solely by the number $N(E,J)$ of open exit channels based
on an analysis of the adiabatic curves. According to statistical
theory, all open channels are assigned equal weight, and the collision
${\bf S}-$matrix elements are given by $\vert
S_{if}(E,J)\vert^2=1/N(E,J)$ for open channels, and zero otherwise. We
note that in the case of reactive (exothermic) systems, the present
approach can be easily extended to include the reactive channels in
the counting of open channels. The cross section is calculated as
\begin{equation}
\sigma_{if}(E)=\frac{\pi}{(2j_1+1)(2j_2+1)k^2}\sum_{J=0}^{\infty} (2J+1)\vert S_{if}(E,J)\vert^2
\end{equation}

The rate coefficients are then computed by integrating the cross
section weighted by a Maxwell-Boltzmann distribution of energies,
\begin{equation}
k_{if}(T) = \Big(\frac{2}{k_BT}\Big)^{3/2} \frac{1}{\sqrt{\pi\mu}} \int_0^\infty E e^{-E/k_BT} \sigma_{if}(E) \ dE \ .
\end{equation}

Note that the present approach allows to save an enormous amount of
both CPU time and memory compared to standard close-coupling
calculations as the adiabatic curves are independent of the
collision energy. Moreover, the convergence of the adiabatic curves with
respect to the ro-vibrational basis is much faster than the
convergence of close-coupling cross sections so that it is not necessary to include closed channels in the basis, even for strongly-bound collisional
systems. We also note that the present approach can be used for
open-shell (polyatomic) molecules showing a complex rotational
structure.
 
\section{Results}

\begin{figure*}
\epsscale{1.1}
\centering
\label{fig_OHpH}
\plotone{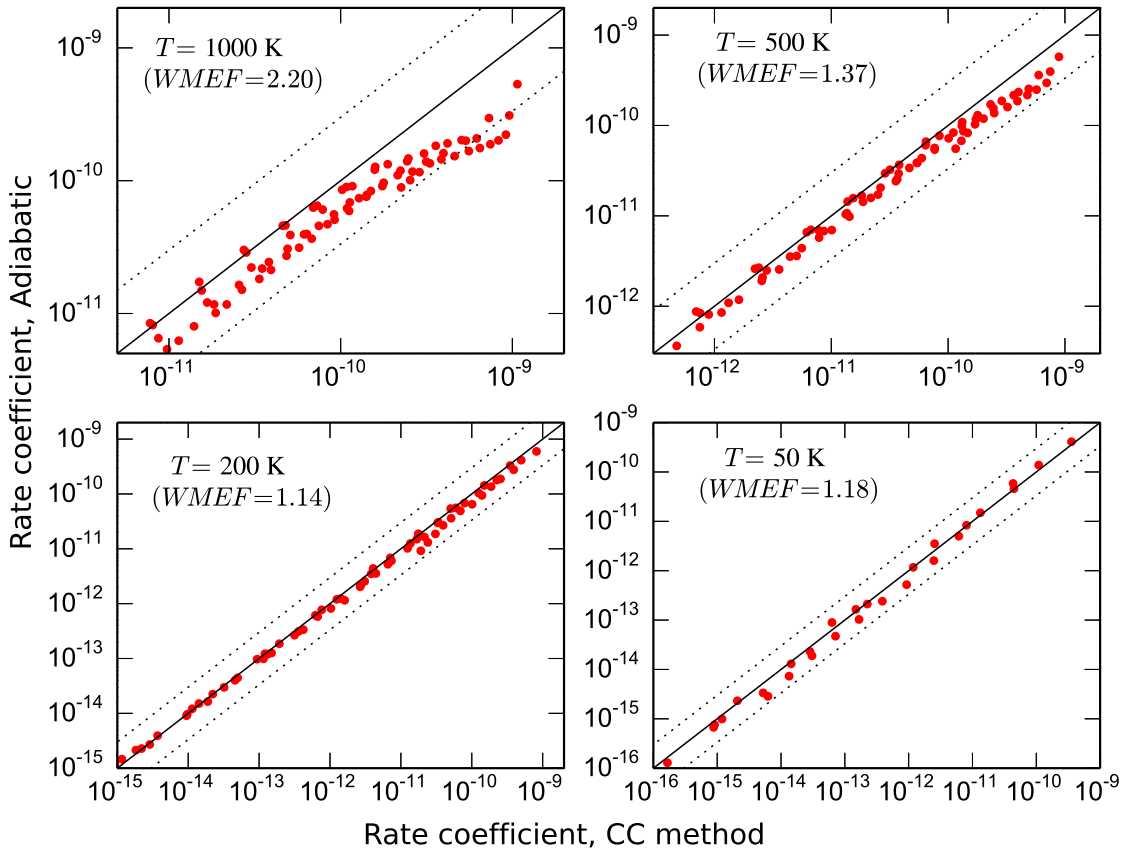}
\caption{Comparison of the rate coefficients (in units of cm$^3$ s$^{-1}$) obtained with the adiabatic model and with the close-coupling (CC) method for OH$^+
  (j_1)$ + H $\rightarrow$ OH$^+(j_1^\prime)$ + H for several
  temperatures. The dashed lines represent a factor of 3 error.}
\end{figure*}

\begin{figure}
\epsscale{1.1}
\centering
\label{fig_propensities}
\plotone{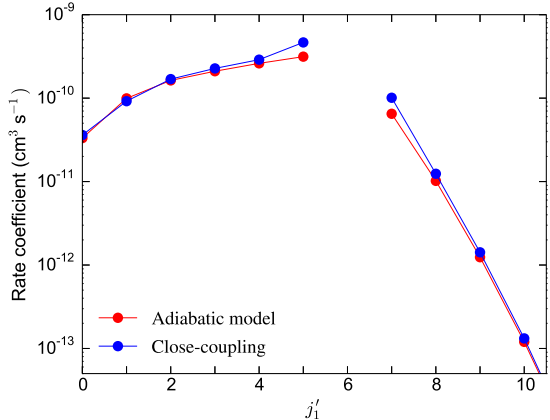}
\caption{Rate coefficients for OH$^+ (j_1=6)$ + H $\rightarrow$
  OH$^+(j_1^\prime)$ + H as function of $j_1^\prime$ at a temperature
  of 200 K.}
    \end{figure}

The method is expected to yield the best results for collisions in
which an intermediate molecular complex can be formed. As an example,
we consider rotationally inelastic OH$^+$--H collisions, for which
accurate close-coupling results are available in the literature
\citep{stoecklin15,Bulut:15}. In this application, we only consider
rigid-rotor collisions on the quartet potential energy surface thus neglecting the (low probability) exchange and reactive channels \citep{Bulut:15}. The intermediate complex, H$_2$O$^+$, has a large
depth (about 0.5 eV or 4,000 cm$^{-1}$) that should favor a
statistical approach. Using the methodology described above, we have
determined the OH$^+$--H rotational rate coefficients in the
temperature range 10--1000~K. Results are presented in
Fig. \ref{fig_OHpH} for four sample temperatures between 50 and 1000~K
and they are compared to the (rigid-rotor) close-coupling results of
\cite{Bulut:15}. At low temperatures (below 300~K), we observe that
the present statistical method predicts the rate coefficients with a
very good accuracy over many orders of magnitude. Up to 300 K, most
rate coefficients are accurate to within 50\%, which is satisfactory for most
astrophysical applications. We note, however, that larger
discrepancies appear as the temperature increases for the dominant transitions. This is expected since at high energies, the collision time
becomes too short to assume formation of a long-lived complex, so that
the hypothesis of an equiprobable distribution of exit channels is not
guaranteed. However, differences remain below a factor of $\sim$3 up
to 1000~K, which is still quite reasonable.  In
Fig. \ref{fig_propensities} we compare the rate coefficients for the
initial state $j_1=6$ at a temperature of 200 K. We observe that the
adiabatic model slightly underestimates the transitions with $\Delta
j_1 =\pm 1$ compared to the close-coupling results, but all other
transitions and propensities are nicely reproduced. The same behaviour
is observed at other temperatures. We note that strong resonance and/or interference effects would necessarily be missed in the statistical approach.

To quantify the discrepancies between the close-coupling results and
the statistical results, we compute the weighted mean error factor
(WMEF) for each temperature, defined as:
\begin{equation}
WMEF=\frac{\sum_i k_i^{\mathrm{CC}}r_i}{\sum_i k_i^{\mathrm{CC}}},
\end{equation}
 where $k_i^{\mathrm{CC}}$ is the rate coefficient for transition $i$
 obtained with the close-coupling method and $r_i$ is the ratio of the
 rate coefficients obtained with the two methods, defined as
 $r_i=\max(k_i^{\mathrm{CC}}/k_i^{\mathrm{adia}},k_i^{\mathrm{adia}}/k_i^{\mathrm{CC}})$
 so that $r_i\geq1$.  We use a weighted average to underline the fact
 that for radiative transfer applications, the largest rate
 coefficients are the most important.
It is clear that the method is very efficient to describe collisions
in strongly-bound systems such as OH$^+$--H with a WMEF of only 1.18
at 50~K. At such low temperature, the approximate wavepacket and semi-
or quasi-classical approaches would fail by much larger factors. The
WMEF increased to 2.20 at 1000~K, which is still reasonable. These
good results confirm that the statistical method is well adapted to
treat systems with ``covalent'' interactions, as expected.

\begin{figure*}
\epsscale{1.15}
\centering
\label{fig_ions}
\plottwo{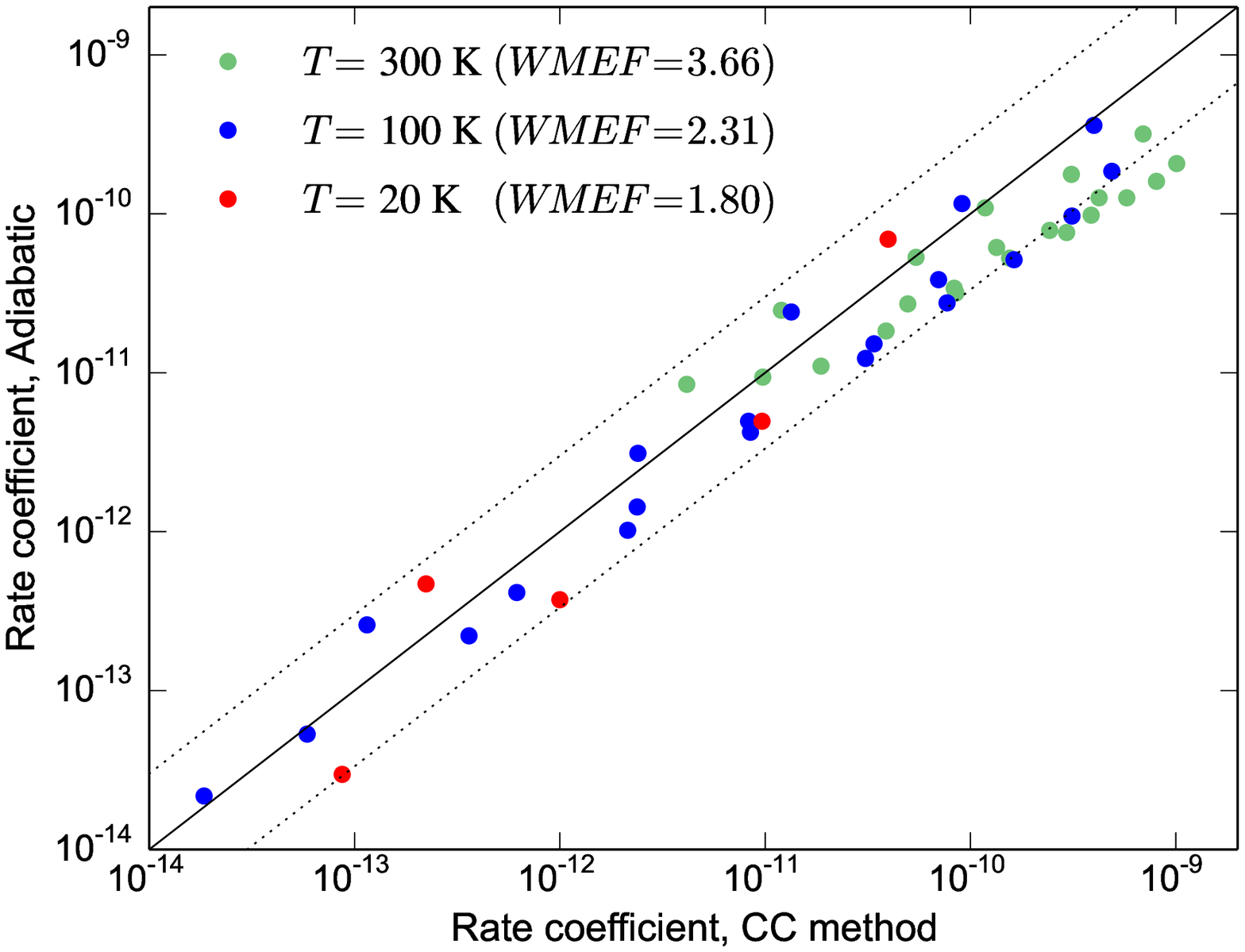}{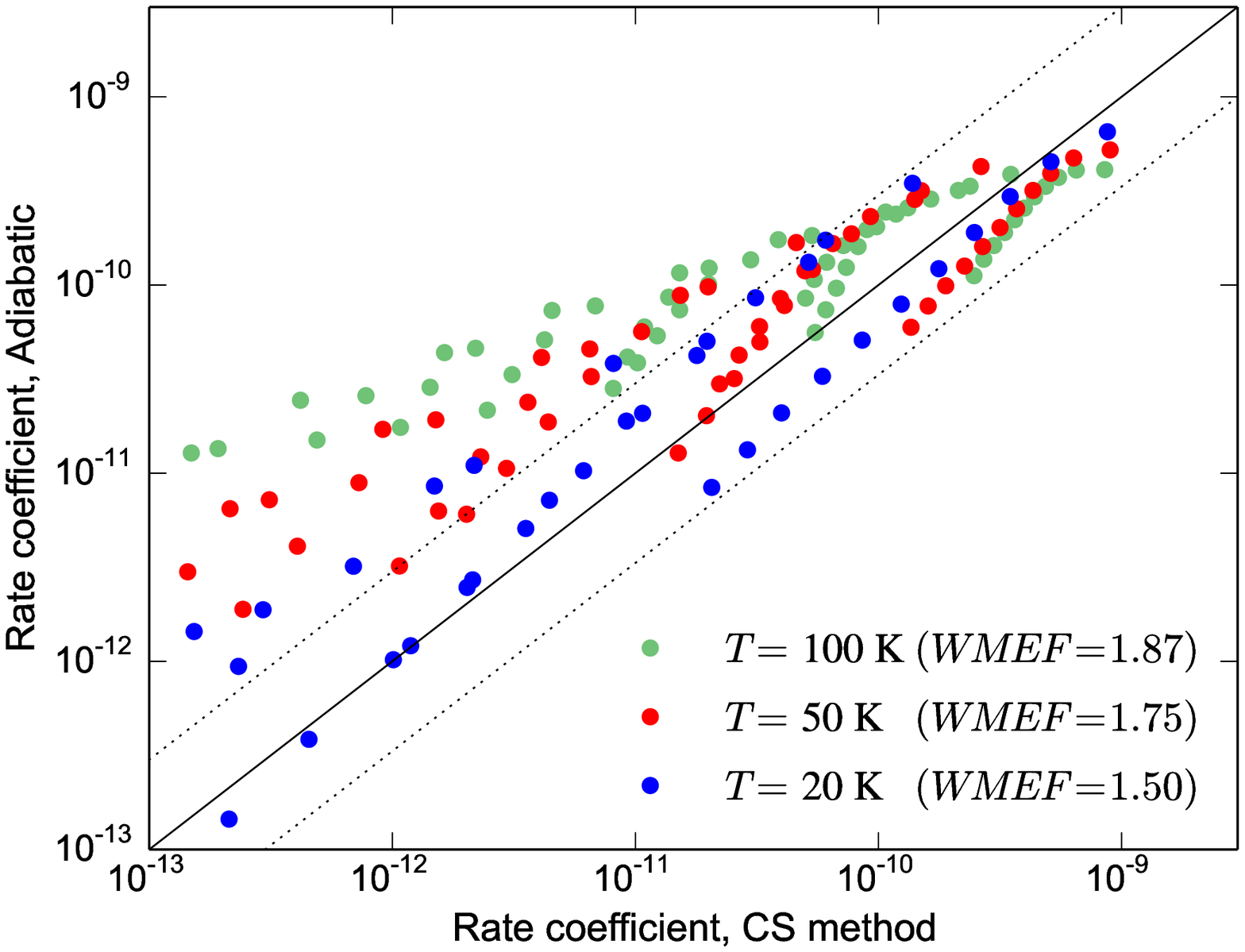}
\caption{Rate coefficients (in units of cm$^3$s$^{-1}$) for OH$^+ (j_1)$ + He $\rightarrow$ OH$^+(j_1^\prime)$ + He (left) and for CN$^-(j_1)$ + H$_2(j_2) \rightarrow$ CN$^-(j_1^\prime)$ + H$_2(j_2)$ (right) for several temperatures. The dashed lines represent a factor of 3 error.}
\end{figure*}

In many astrophysical environments, such as the cold interstellar
medium, the most abundant colliding partners are not hydrogen atoms
but H$_2$ and He. In this case, the corresponding intermediate
complexes have much shallower well depth, and one might wonder whether
the statistical method is still applicable. We first consider
collisions of molecular ions with He or H$_2$. These collisions
provide an intermediate case as a non-covalent but strong van der
Walls interaction occurs due to the charge-induced dipole potential.
Fig.~\ref{fig_ions} illustrates the results for a cationic system,
OH$^+$--He, and an anionic system, CN$^-$--H$_2$. The well depths of
the potential energy surfaces are very similar, 730~cm$^{-1}$ and 750~cm$^{-1}$
respectively. The rotational rate coefficients have been calculated
with the close-coupling method for OH$^+$--He \citep{Gomez:14} and the
coupled-states method for CN$^-$--H$_2$ \citep{Klos2011}. For this
latter system, close-coupling calculations have shown that the
coupled-states approach is accurate to better than 50\%. In the case
of OH$^+$--He, we observe a reasonable agreement (better than a factor
3 for most transitions) for the three temperatures considered,
although at 300~K the dominant rate coefficients are not well
reproduced.
For CN$^-$--H$_2$, the agreement is not that good, which might reflect
the fact that H$_2$ has a large rotational constant and does not
behave statistically.
This is particularly the case for transitions with small rate
coefficients ($<10^{-11}$ cm$^3$s$^{-1}$). On the other hand, the
dominant transitions are rather accurately reproduced at all
temperatures. As for OH$^+$--H collisions, we note that the agreement
between the statistical and the coupled-channel results decreases with
increasing temperature. We conclude that the statistical approach can
be useful also for (non-covalent) strongly bound van der Waals
systems, provided that low temperatures are considered.

Finally, we consider collisions in which the intermediate complex is
bound by weak van der Waals forces. Due to its large abundance in many
astrophysical environments, we used CO in collision with He and H$_2$,
for which close-coupling rate coefficients are available
\citep{cecchi2002,yang2010}. In both cases the depth of the potential
energy surface is very small, 24 and 94~cm$^{-1}$, respectively. Quite
unexpectedly, the statistical method still performs relatively well,
with most dominant transitions (i.e. rate coefficients above
10$^{-11}$cm$^3$s$^{-1}$) being reproduced with an error less than a
factor of 3 and WMEF in the range 1.5--3. On the other hand, the
scatter of points is much larger than for the previous systems, as
expected from the much smaller well depths. In addition, for CO--H$_2$ the WMEF decreases with {\it increasing} temperature, which might suggest the method is out of its applicability domain.

\begin{figure*}
\epsscale{1.15}
\centering
\label{fig_VdW}
\plottwo{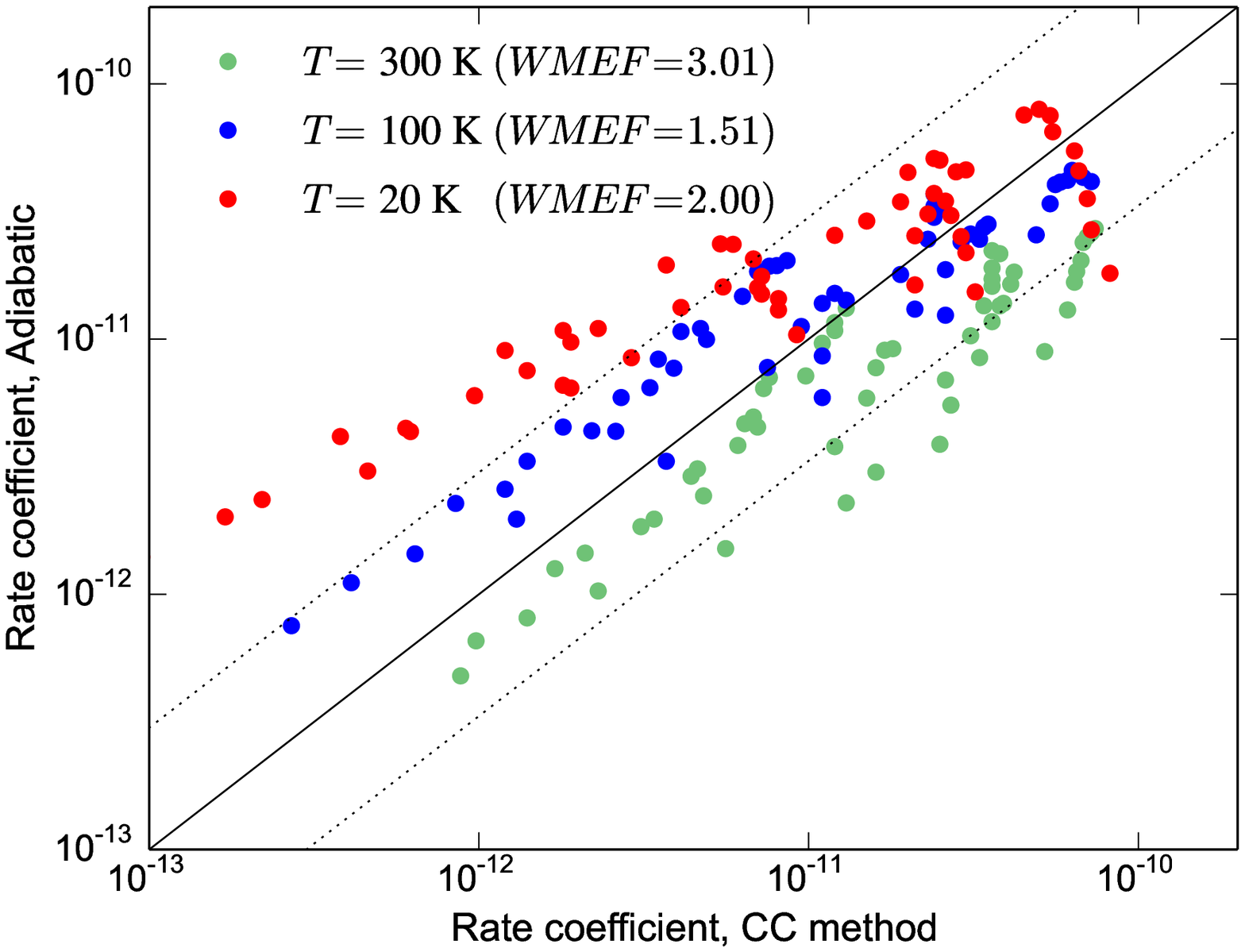}{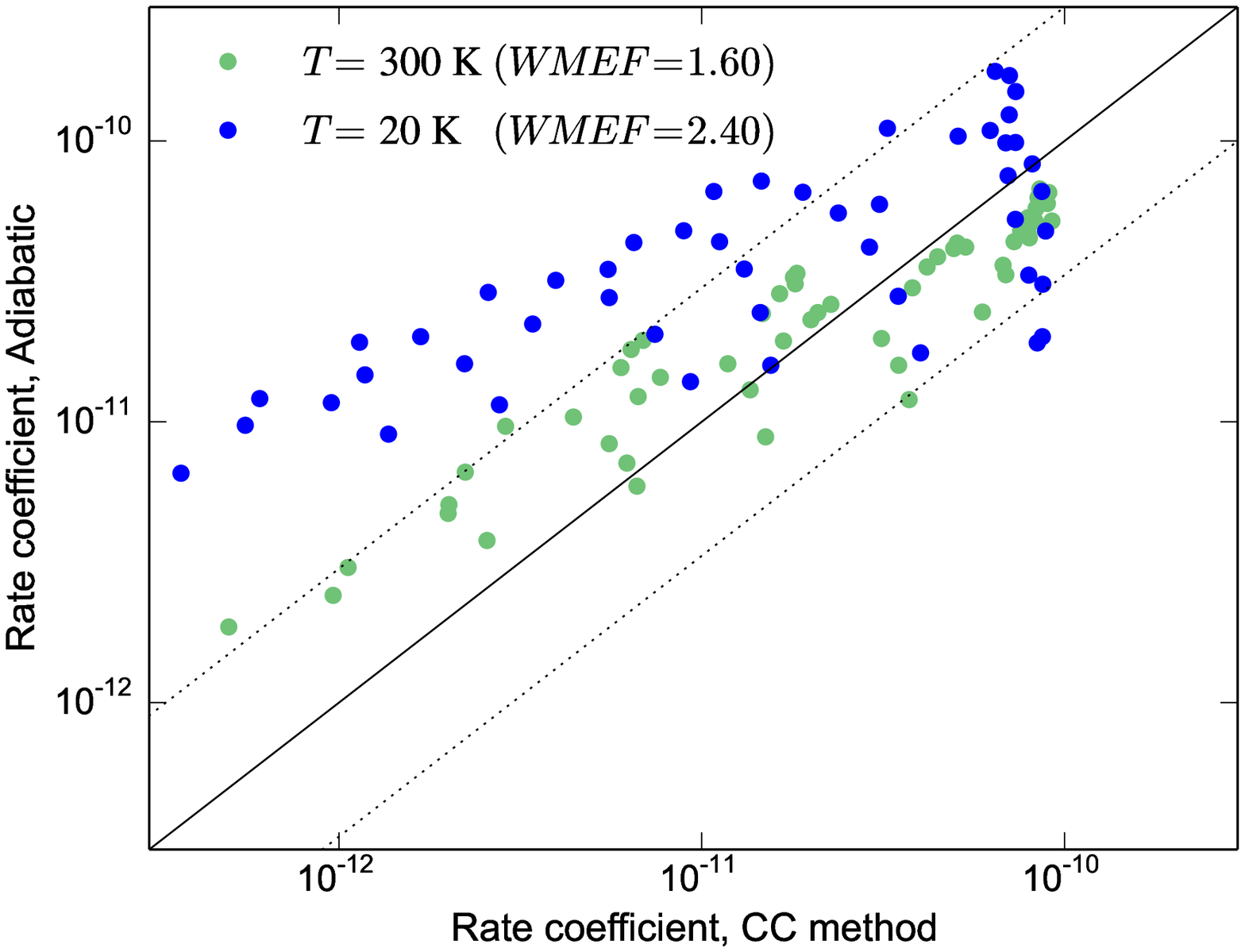}
\caption{Rate coefficients  (in units of cm$^3$s$^{-1}$) for CO$(j_1)$ + He $\rightarrow$ CO$(j_1^\prime)$ + He (left) and for CO$(j_1)$ + H$_2(j_2) \rightarrow$ CO$(j_1^\prime)$ + H$_2(j_2)$ (right) for several temperatures. The dashed lines represent a factor of 3 error.}
\end{figure*}

In summary, the statistical method is found to provide average
accuracies better than 50\% for strongly-bound (covalent) collisional
systems up to room temperature. At higher temperature, or for systems
with weaker interactions, larger inaccuracies are observed, as
expected from the basic assumption of the method.

\section{Discussion and conclusions}

In this letter, we have presented a new statistical method to compute
inelastic rate coefficients for interstellar molecules.  The method
has been benchmarked versus ``exact'' coupled-channel calculations for
five collisional systems (OH$^+$--H, OH$^+$--He, CN$^-$--H$_2$, CO--He
and CO--H$_2$). These five systems include a strongly-bound molecular
system, two strongly-bound van der Waals systems and two weakly-bound
systems. For the strongly-bound case, we found the statistical results
to be in excellent agreement with the coupled-channel results,
especially at low temperatures where the lifetime of the collision
complex is the longest. When the temperature increases above room
temperature, the agreement decreases as expected. For the two
strongly-bound van der Waals systems, the agreement is still (perhaps
surprisingly) rather satisfactory. For the two weakly-bound systems,
the accuracy decreases significantly, as could be anticipated.

Whereas accurate collisional rate coefficients are now available for
many non-reactive molecules \citep{roueff:13}, very few such reliable
data exist for inelastic collisions involving reactive radicals such
as NH or reactive ions such as H$_3^+$, H$_2$O$^+$ and H$_3$O$^+$, for which reactive and inelastic channels are competing. In order to
provide the astrophysical community with collisional data for these
systems, the helium atom is generally employed as a substitute for H
or H$_2$ for simplicity. However, it is now well established that
significant differences exist between He, H and H$_2$ collisional rate
coefficients \citep{roueff:13}. Collisions with He atoms can thus
provide guidance at the order-of-magnitude level but they can
certainly not reproduce H or H$_2$ to within 50\% accuracy. 
The present approach leads to much more accurate results for inelastic collisions and should be considered as a useful alternative to prohibitive close-coupling calculations. Future applications will include exothermic reactive collisions such as the triatomic CH$^+$+H system for which full-dimensional close-coupling calculations are available \citep{Werfelli2015}.

Finally, we note that in the coma of comets and in protoplanetary
disks, new partners enter the collision game. Thus, in comets,
molecules are excited by H$_2$O which is the most abundant neutral
species in the coma. The computation of rates for the excitation of molecules due to
H$_2$O collisions is not yet numerically accessible through quantum
coupled-channel calculations due to the large potential wells
\citep{sanchez17,Kalugina:18}. The present approach has no such CPU
limit and as soon as a potential energy surface is available, collisional data can be
easily predicted. As a result, the present statistical method will
allow to study the collisional excitation of cometary molecules such
as CO, H$_2$O or HCN by H$_2$O and to provide the first collisional
rate coefficients for H$_2$O--molecule systems.

\section{Acknowledgments}

We thank T. Stoecklin, M. Alexander, and P. Dagdigian for useful discussions.
This research was supported by the French National Research Agency
(ANR) through a grant to the Anion Cos Chem project
(ANR-14-CE33-0013). We also acknowledge the financial support of the
Region Normandie through the Ç Actions Th\'ematiques Strat\'egiques È
project managed by Normandie University.

\end{document}